\documentclass[prc,nofootinbib,onecolumn,showpacs,showkeys]{revtex4}
\usepackage[dvips]{graphicx}

\newcommand{\bea}{\begin{eqnarray}}
\newcommand{\eea}{\end{eqnarray}}
\newcommand{\w}{$\omega$ }
\newcommand{\bc}{\begin{center}}
\newcommand{\ec}{\end{center}}
\newcommand{\bfi}{\begin{figure}}
\newcommand{\efi}{\end{figure}}

\begin{document}

\title{Inclusive \w photoproduction off nuclei}
\author{P. M\"uhlich}
\email{pascal.muehlich@theo.physik.uni-giessen.de}
\author{T. Falter}
\author{U. Mosel}
\affiliation{Institut f\"ur Theoretische Physik, Universit\"at Giessen, D--35392 Giessen, Germany}

\begin{abstract}
We investigate inclusive \w photoproduction off complex nuclei, concentrating on the feasibility to examine a possible in-medium change of the \w meson properties by observing the $\pi^0\gamma$ invariant mass spectrum. The simulations are performed by means of a BUU transport model including a full coupled-channel treatment of the final state interactions. In-medium changes of the \w spectral density are found to yield a moderate modification of the observables as compared to the situation in free space. Also the effects of a momentum dependence of the strong \w potential are discussed.
\end{abstract}

\pacs{25.20.-x, 13.60.-r, 13.60.Le, 14.40.-n}
\keywords{photonuclear reactions, photon interactions with hadrons, meson production, mesons}

\maketitle

\section{Introduction}

The expectation to gather experimental evidence for the partial restoration of chiral symmetry at finite baryon density and temperature has caused great interest in the properties of the light mesons in a dense and hot environment \cite{Asakawa:1993,Rapp:2000,Post,Cabrera:2002}. Various theoretical predictions indicate that the observation of an in-medium modification of the vector meson masses can provide a unique measure to what degree chiral symmetry is broken in the strongly interacting medium \cite{Hatsuda}. For instance, the scaling law proposed by Brown and Rho indicates a direct locking of the vector meson masses and the chiral condensate, yielding a considerable decrease of the in-medium $\rho$ and \w meson masses already at normal nuclear matter density ($0.17~\mathrm{fm}^{-3}$) \cite{Brown:1991}. On the other hand, a careful analysis of QCD sum rules (QCDSR) has shown that there is no direct, QCD-mandated connection between hadron masses and quark condensates \cite{Leupold:1998QCDSR}. An outcome of this study was that the QCDSR could also be fulfilled by increasing the width of the hadron in medium. In both scenarios, however, there is consensus that hadronic strength is shifted downwards to lower masses.

Moreover, various experimental observations have been addressed to a modification of the $\rho$ resonance in the nuclear medium \cite{Brown:1990,Rapp:1999}. For instance, the accumulation of spectral strength in the $e^+e^-$ mass spectrum below the $\rho$ meson pole measured in heavy ion reactions \cite{CERES:1995,CERES:2002} can be understood in terms of a broadening of the $\rho$ spectral function and/or a reduction of the $\rho$ pole mass in the nuclear medium \cite{Li:1995,Cassing:1997,Cassing:1999}. Attributing such observations to conventional nuclear physics effects or to more pioneering phenomena is presently the great challenge to  nuclear theory. Connecting the outcome of theoretical models concerning the intrinsic properties of hadrons to experimental observables is an important issue which will be addressed in the present paper.

Understanding the dependence of the vector meson properties in a quantitative way is still an outstanding problem. Predictions for the values of the $\rho$ mass at normal nuclear matter density range from 530 MeV to 730 MeV \cite{Asakawa:1993,Hatsuda,Koike:1995,Jin:1995,Saito:1997,Leupold:1998} compared to its vacuum mass of 769 MeV. Also the predicted values for the in-medium mass of the \w meson are spread over the wide range from 640 MeV to 765 MeV \cite{Klingl:1997,Friman:1998,Saito:1998,Saito:1999,Klingl:1998} (782 MeV in vacuum). Furthermore the predicted values for the broadening of the $\rho$ (up to 300 MeV) \cite{Klingl:1997,weidmann,Effenberger:1998,Cabrera:2000} and $\omega$ (up to 50 MeV) \cite{Klingl:1997,Friman:1998} mesons at rest cover a large set of possible hadronic scenarios. This emphasizes the importance to make an experimental determination of the $\rho$ and \w spectral distributions in the nuclear medium.

In this respect, the cleanest way to investigate the vector meson properties in nuclear matter is provided by exploiting the photoproduction of these mesons from complex nuclei \cite{mosel:1997}. As compared to heavy ion collisions, where temperature and density vary dramatically with time, the nucleus stays more or less intact in a photonuclear reaction. Moreover, in contrast to hadron induced reactions photoproduction has the advantage, that due to its electromagnetic coupling to the nucleons the reaction probability of the photon is almost the same for all nucleons inside the nucleus. Therefore, all densities of the static density distribution provided by the target nucleus are probed, in principle giving rise to large density effects on the measured observables.

A drawback of exploiting any hadronic decay mode is the distortion of the final state by elastic, inelastic and absorptive final state interactions (FSI). Up to the present the only way to study such effects in a quantitative way is the utilization of semi-classical transport models. In Ref. \cite{Muehlich:2002}, we presented a similar study of the reaction $\gamma A\rightarrow\phi X\rightarrow K^+K^-X'$ within a Boltzmann-Uehling-Uhlenbeck (BUU) transport model including a full coupled-channel treatment of the FSI. As an outcome of this work we found that the in-medium modifications of the $\phi$ meson are not visible in the $K^+K^-$ mass spectrum due to the strong FSI and the additional effects of the strong and electromagnetic potentials of the final state particles.

Dilepton spectroscopy in principle provides a tool to measure the vector meson in-medium spectral densities without distortion due to FSI \cite{Effenberger:1998}. Such an experiment is presently being analyzed at JLAB \cite{Weygand}. Another promising experiment, where there is only one strongly interacting particle in the exit channel is the measurement of the $\pi^0\gamma$ invariant mass spectrum. In the present paper we study inclusive \w photoproduction off p, $^{12}$C and $^{92}$Nb at photon energies of 1.2 GeV and 2.5 GeV, which is the kinematical regime of the TAPS \cite{Gabler:1994}/Crystal Barrel \cite{Aker:1992} experiment at ELSA.

As already pointed out in our previous work, a great advantage of our model is the inclusion of inclusive elementary photoproduction processes. These processes are in particular important when the issue of medium modifications is addressed, since \w mesons from inclusive reactions ($\gamma N\rightarrow \omega X, X\neq N$) are produced with lower momenta as compared to \w mesons from exclusive reactions ($\gamma N\rightarrow \omega N$). Hence, \w mesons from inclusive elementary reactions decay with higher probability inside the nucleus and, therefore, carry information about the in-medium properties of the decayed resonances. In our approach the inclusive photoproduction processes are described by using vector meson dominance (VMD) and the event generator FRITIOF \cite{FRITIOF} which is also used to describe high energy collisions during the FSI.

Most of the currently available model predictions for the \w meson mass and width inside nuclei are concerned with \w mesons at rest \cite{Klingl:1997,Friman:1998,Saito:1998,Saito:1999,Klingl:1998} (see, however \cite{Leupold:1998QCDSR}). Indeed \w mesons originating from photonuclear reactions are produced with finite momenta, which already at beam energies slightly above threshold can achieve quite large values. Using an estimate on the real part of the strong \w meson potential obtained by a subtracted dispersion relation we also study the in medium behavior of such fast moving \w mesons. In this respect the measurement of the \w spectral density at different beam energies could also shed some further light on the momentum dependence of the \w mass in nuclear matter.

Photoproduction of \w mesons off nuclei has already been studied in Ref. \cite{Cassing:2001}. However, the model of Ref. \cite{Cassing:2001} can only be applied to energies slightly above threshold, since this model involves only the exclusive photoproduction process. Moreover, the authors of Ref. \cite{Cassing:2001} used a momentum-independent in-medium width and also did not study a possible momentum-dependence of the nuclear \w potential. Furthermore, they did not separate the effects of collisional broadening and a downward shift of the \w mass. On the other hand, they did concentrate on effects of the detector resolution as well as on the determination and suppression of the background from interfering reactions which will also be useful to the present work.

Our paper is organized as follows. In section \ref{photonnucleon} we explain how we describe the photon-nucleon interaction. The transport model itself is sketched in section \ref{buu}. In section \ref{results} we give a detailed discussion of our results. We close with a short summary in section \ref{summary}.

\section{Model}\label{model}

\subsection{Primary photon-nucleon interaction}\label{photonnucleon}

In the impulse approximation the photon-nucleus interaction is reduced to the interaction of the photon with a single bound nucleon of the target nucleus. As explained in more detail in Ref. \cite{Falter:2002}, we use the event generator FRITIOF to simulate particle production in high energy photon-nucleon reactions. Here the idea that the photon might fluctuate into a virtual vector meson $V=\rho^0,\omega,\phi$ (vector meson dominance) is used to replace the incoming photon by a $\rho$, $\omega$ or $\phi$ meson:
\bea
|\gamma\rangle\rightarrow\sum\limits_{\rho,\omega,\phi}\frac{e}{g_V}|V\rangle.
\eea
As we have already pointed out in Ref. \cite{Muehlich:2002} one cannot account properly for the processes $\gamma N\rightarrow V N$ and $\gamma N\rightarrow V\Delta$ within the FRITIOF model and, therefore, has to parameterize these channels explicitly.

\begin{figure}
\begin{center}
\includegraphics[scale=1.2]{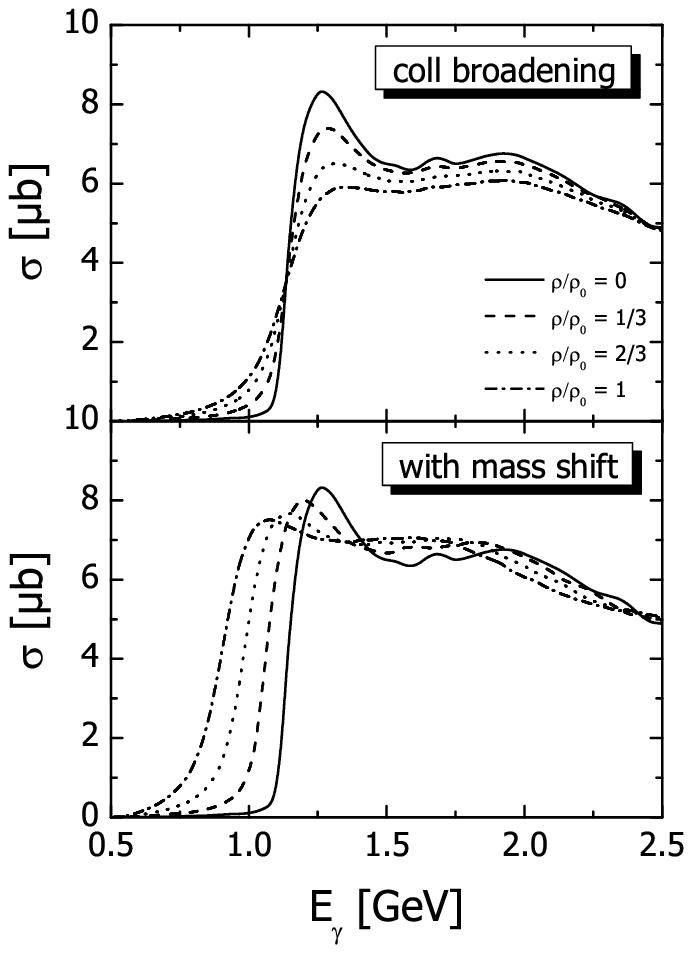}
\caption{Total cross section for exclusive \w photoproduction from a (bound) proton at several nucleon densities. For the upper panel a collisional broadening in accordance with the low-density theorem has been considered, whereas the lower panel includes an additional downward shift of the \w mass. The case $\rho=0$ describes the free cross section on the proton. Data points are shown in Fig. \ref{figure03}.}
\label{figure01}
\end{center}
\end{figure}

For the exclusive \w photoproduction process we make the following ansatz:
\bea
\sigma_{\gamma N\rightarrow\omega N}(s,\rho_N(\vec r))=\frac{1}{p_i s}\int d\mu\mathcal{S}_{\omega}(\mu,\rho_N(\vec r))p_f(\mu)|\mathcal{M}_{\omega}(Q^2(\mu))|^2,
\eea
where $p_i$ is the initial cm-momentum of the $\gamma N$ system, $p_f$ is the final cm-momentum of the $\omega N$ system and $\mathcal{S}_{\omega}$ is the \w spectral function. The squared matrix element $|\mathcal{M}_{\omega}|^2$ is obtained from the experimental photoproduction cross section in vacuum by assuming the matrix element to be independent of the \w mass $\mu$:
\bea
|\mathcal{M}_{\omega}(s)|^2=\sigma^{\mathrm{exp}}_{\gamma N\rightarrow\omega N}(s){p_i s}   \left/ {\int d\mu \mathcal{S}_{\omega}(\mu,\rho_N=0)p_f(\mu)}\right..
\eea
The production cross section $\sigma^{\mathrm{exp}}_{\gamma N\rightarrow\omega N}$ has been measured with the SAPHIR detector at ELSA \cite{SAPHIR} (see also Fig. \ref{figure03}).
In order to describe the photoproduction of \w mesons with masses $\mu$ below the pole mass $m_{\omega}^0$ at finite nuclear densities we follow Ref. \cite{Larionov:2003} and extend this matrix element to subthreshold energies by defining a new invariant, namely the 'free energy' 
\bea
Q(\mu)=\sqrt{s_{\mathrm{th}}(m_{\omega}^0)}-\sqrt{s_{\mathrm{th}}(\mu)}+\sqrt{s}.
\eea
Here $\sqrt{s_{\mathrm{th}}(m_{\omega}^0)}$ denotes the threshold energy for the production of \w mesons of mass $m_{\omega}^0=0.782$ GeV and $\sqrt{s_{\mathrm{th}}(\mu)}$ is the threshold energy for the photoproduction of \w mesons of mass $\mu$. In Fig. \ref{figure01} we show the resulting cross section at four different densities. In the upper panel we considered an in-medium broadening of the \w according to the low density theorem:
\bea
\label{width}
\Gamma_{\mathrm{coll}}(\rho_N,p_{\omega})=\gamma\rho_N \int\limits^{p_F(\rho)} d^3p_N~ v_{\omega N} \sigma_{\omega N},
\eea
where $\gamma$ is the Lorentz factor for the transformation from the nuclear rest frame to the \w rest frame, $p_F$ is the local Fermi momentum, $v_{\omega N}$ denotes the \w nucleon relative velocity and $\sigma_{\omega N}$ is the \w nucleon total cross section. Using the cross section parameterization of Ref. \cite{Lykasov:1998} that we are employing for $\omega N$ collisions throughout the FSI this leads to an additional collision broadening of 40 MeV at normal nuclear density and vanishing \w momentum.
For the lower panel of Fig. \ref{figure01} we considered an additional \w mass shift according to Ref. \cite{Hatsuda}:
\bea
\label{shift}
m_{\omega}^*=m_{\omega}^0\left(1-0.16~\frac{\rho_N}{\rho^0}\right),
\eea
which leads to an obvious lowering of the \w photoproduction threshold at finite density. There is an additional effect on the threshold behavior due to a broadening of the nucleon spectral function due to short range correlations \cite{Juergen}. This effect is, however, much less important than that shown in Fig. \ref{figure01} and we will, therefore, disregard it in the following.

For the angular distribution we use a Regge parameterization according to Ref. \cite{Donnachie:1999} plus an additional contribution which is constant in the four momentum transfer $t$ in accordance with the experimental data from Ref. \cite{SAPHIR}.

\begin{figure}
\begin{center}
\includegraphics[scale=1.2]{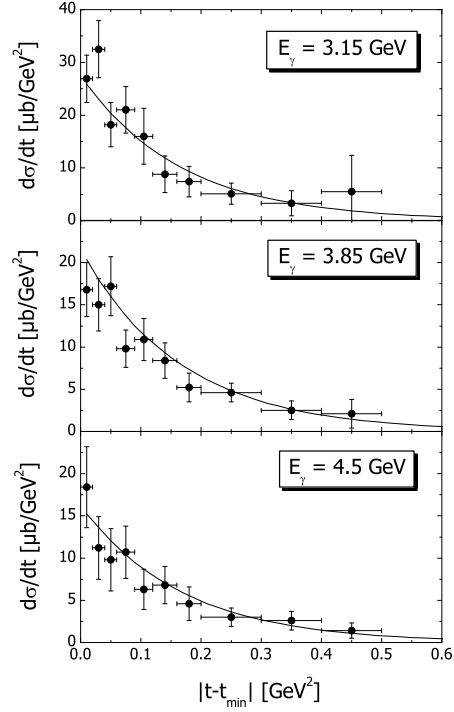}
\caption{Differential cross section for the process $\gamma p\rightarrow\omega\Delta^+$ together with experimental data from Ref. \cite{Barber:1984}.}
\label{figure02}
\end{center}
\end{figure}

\begin{figure}
\begin{center}
\includegraphics[scale=1]{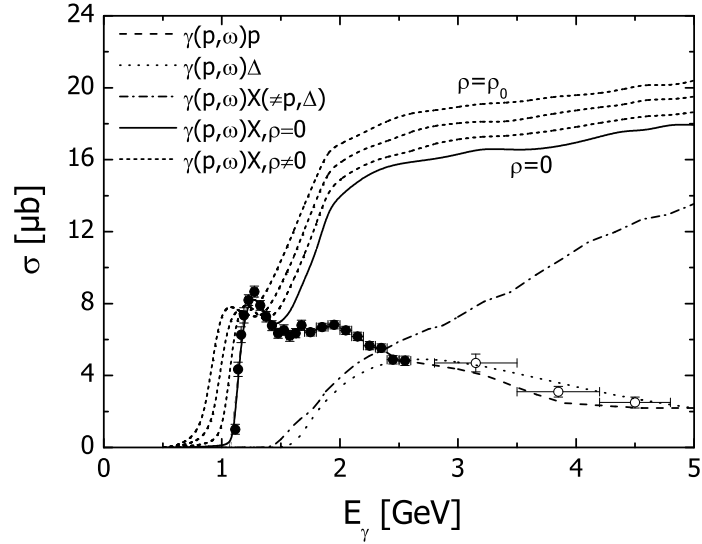}
\caption{Total cross section for \w photoproduction off the proton decomposed in the several contributing reaction channels (see legend). The experimental data for the exclusive process stem from Ref. \cite{SAPHIR} (solid circles) and the data for the process $\gamma p\to\omega\Delta^+$ are taken from Ref. \cite{Barber:1984} (open circles). In addition, the total cross section $\sigma(\gamma p\to\omega X)$ at nuclear densities $\rho=0, \rho_0/3, 2 \rho_0/3, \rho_0$ is shown.}
\label{figure03}
\end{center}
\end{figure}

For the process $\gamma N\rightarrow \omega\Delta$ we use the following parametrization of the total cross section:
\bea
\sigma_{\gamma N\rightarrow\omega\Delta}=\frac{1}{p_i s}\int d\mu_{\Delta}\mathcal{S}_{\Delta}(\mu_{\Delta})p_f(\mu_{\Delta}) \times\frac{A}{(\sqrt{s}-M)^2+\Gamma^2/4},
\eea
where the constants $A,M$ and $\Gamma$ are fitted to the experimental cross section of Ref. \cite{Barber:1984}, yielding $A=47.3~\mu \mathrm{b}~\mathrm{GeV}^2$, $M=2.3~\mathrm{GeV}$ and $\Gamma=1.8~\mathrm{GeV}$. The angular distribution is parameterized as follows:
\bea
\frac{d\sigma_{\gamma N\rightarrow\omega\Delta}}{dt}\propto\exp{(Bt)},
\eea
where $B=6.0~\mathrm{GeV}^{-2}$, compare Fig. \ref{figure02}. For both the exclusive process $\gamma N\rightarrow\omega N$ and $\gamma N\rightarrow\omega\Delta$ we assume the cross section on the neutron to be the same as on the proton.

In Fig. \ref{figure03} we show the total \w photoproduction cross section off the proton decomposed into the several reaction channels. At $E_{\gamma}=1.37$ GeV the phase-space for inclusive \w photoproduction opens up, namely the production of the $\omega(782)\pi N$ final state becomes energetically possible. At an incident beam energy of 2.5 GeV, which we are also going to consider in the following, the channel $\gamma N\to\omega\Delta$ and the inclusive processes $\gamma N\to\omega X(\ne N,\Delta)$, that we describe employing the event generator FRITIOF, become as important as the exclusive \w production channel. Even more relevant in the present context are the differing momentum distributions of the exclusive and inclusive reaction mechanisms. This issue we are going to discuss along with the results of our calculations. In addition, we show the total inclusive \w photoproduction cross section from a nucleon at various nuclear densities. Here the broadening and shift of the \w mass distribution according to Eqs. \ref{width} and \ref{shift} have been considered, leading to an increasing phase-space for \w production at finite densities.

\subsection{Transport model}\label{buu}

The calculations are performed using the transport model previously applied to $K^+K^-$ \cite{Muehlich:2002} and dilepton \cite{Effenberger:1998} photoproduction off nuclei. A more extensive description of the model has also been given in Refs. \cite{Teis:1997,Effenberger:1997,Effenberger:1999,Effenberger:2000,Lehr:2000}. The propagation of the final state of the elementary photon-nucleon interaction is treated by the semi-classical BUU equation:
\bea
\left(\partial_{t}+\partial_{\vec p}\mathcal{H}\partial_{\vec r}-\partial_{\vec r}\mathcal{H}\partial_{\vec p}\right)F_i(\vec r,\vec p,t,\mu)=I_{\mathrm{coll}}[F_1,...,F_i,...,F_N],
\label{BUUEQN}
\eea
which describes the time evolution of the spectral phase space density $F_i(\vec r,\vec p,t,\mu)$ of particles of type $i$ that can interact via binary reactions. Besides the mesons $(\pi,\eta,\rho,\omega,\phi,...)$ and the nucleons the model also contains the baryonic resonances which can be produced either in the photon-nucleon interaction or during the FSI. For baryons the Hamiltonian $\mathcal{H}$ contains a mean-field potential which depends on the position and momentum of the particle. Eq. (\ref{BUUEQN}) allows for an off-shell transport of collision-broadened particles which is essential for the present case where the \w meson picks up a considerable collision width.

The collision term on the right hand side of the BUU equation accounts for the creation and annihilation of particles as well as for elastic and inelastic scattering from one position in phase space to another, including Pauli blocking factors for outgoing nucleons. The BUU equations for different particle species are coupled through the collision term and the mean-field. The set of coupled equations is solved by a test particle ansatz for the spectral phase space densities. The modification of the vector meson in-medium properties according to equations (\ref{width}) and (\ref{shift}) is introduced by means of a scalar off-shell potential proportional to the density in line with the low-density theorem. This potential consistently guarantees that those particles which are broadened and/or shifted in the nuclear medium regain their vacuum spectral function as they propagate out of the nucleus. For further details, see Refs. \cite{Muehlich:2002,Effenberger:1998}.

\section{Results}\label{results}

Before we turn to the discussion of medium modifications of the \w meson, we discuss some standard nuclear effects. Going from the proton to finite nuclear targets the photoproduction cross section involves the effects of Fermi motion, Pauli blocking, the binding energy of the nucleons, and interactions of the produced final state with the surrounding nuclear medium (FSI). Furthermore, also the effect of nuclear shadowing has been included in our calculations. Shadowing arises from the absorption of the hadronic components of the photon, which in the simple vector meson dominance picture consist of the light vector mesons $V=\rho^0,\omega,\phi$, on their way through the nucleus. This effect yields a reduction of the interaction probability of the photon with nucleons on the backside of the nucleus with respect to the incoming photon direction and, hence, to a decrease of the total nuclear cross section as compared to $A$ times the nucleonic cross section. For details, see Ref. \cite{Shadowing}. 

\begin{figure}
\begin{center}
\includegraphics[scale=1.4]{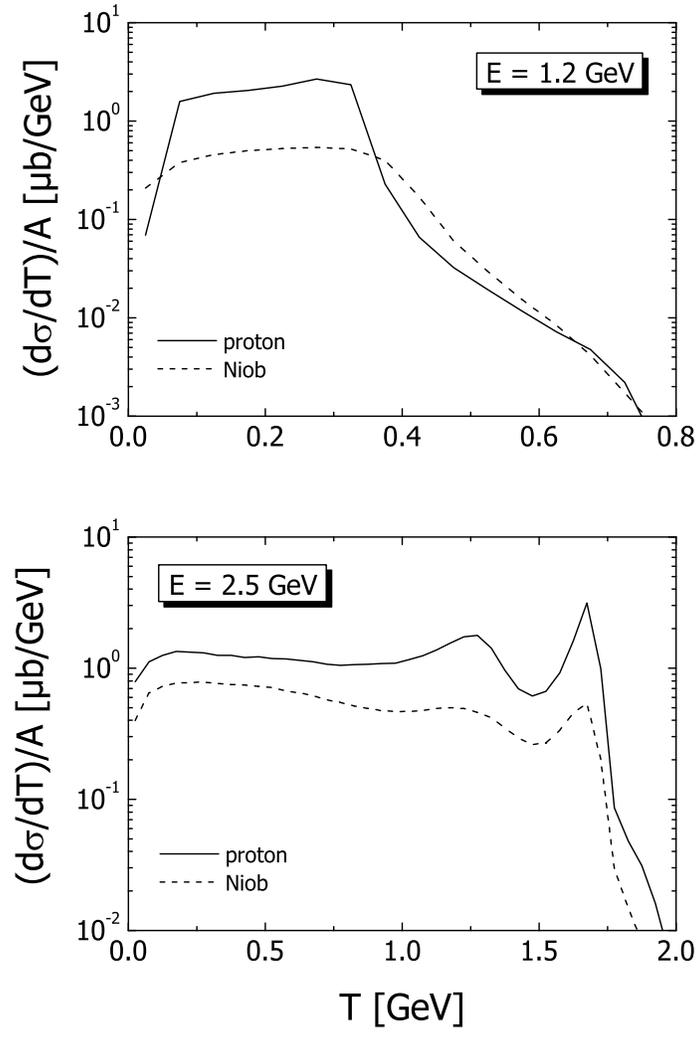}
\caption{Kinetic energy distribution for inclusive \w photoproduction at 1.2 and 2.5 GeV incident photon energy. Comparison of proton and Niobium targets.}
\label{figure04}
\end{center}
\end{figure}

In Fig. \ref{figure04} we compare the kinetic energy spectrum for photoproduction of $\pi^0\gamma$ pairs off the proton and $^{92}$Nb at two different energies of the incident photon beam, 1.2 GeV and 2.5 GeV. The structure that is visible in the proton cross section at 1.2 GeV simply reflects the kinematically allowed region for \w mesons that stem from the process $\gamma N\rightarrow\omega N$ and have a mass equal to the pole mass 782 MeV. The high energy tail is caused by the small but finite \w width. At 2.5 GeV the $\gamma N\rightarrow\omega\Delta$ production mechanism becomes important, leading to a second peak at $\simeq 1.25$ GeV in the proton cross section. By looking at our results for the $^{92}$Nb target one can observe a smearing-out and a reduction of the differential cross section. While at 1.2 GeV these effects are solely due to Fermi motion and the FSI, part of the reduction at 2.5 GeV is caused by the initial state interactions of the photon (shadowing). It is also interesting that in the case of 2.5 GeV incident beam energy the cross section at large \w momenta is lowered by nearly one order of magnitude, whereas at low momenta one observes only a reduction of about a factor of two. This can be attributed to a slowing down of \w mesons due to elastic and inelastic FSI, which we found to be even more important for the photoproduction of $\phi$ mesons in nuclei, see Ref. \cite{Muehlich:2002}. 

\begin{figure}
\begin{center}
\includegraphics[scale=1.4]{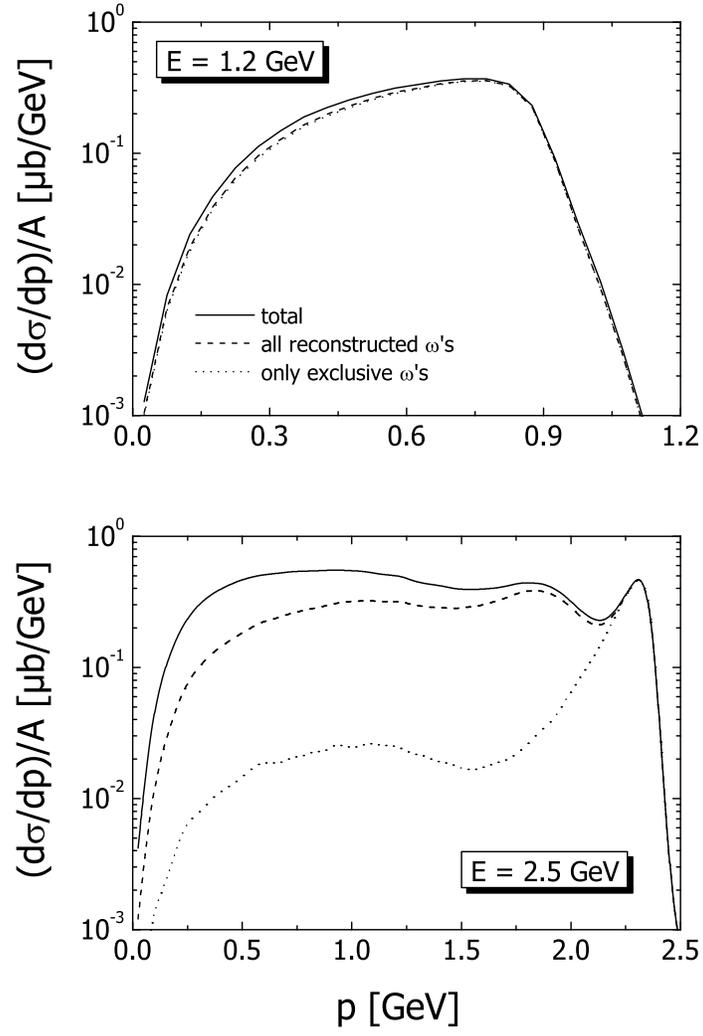}
\caption{Momentum distribution of reconstructed \w mesons in the reaction $\gamma Nb\rightarrow\pi^0\gamma X$ for 1.2 and 2.5 GeV incident photon energy. The solid curve contains all reconstructed $\pi^0\gamma$ pairs without any mass cut.}
\label{figure05}
\end{center}
\end{figure}

In Fig. \ref{figure05} we show the momentum differential cross section for $\pi^0\gamma$ photoproduction off Nb. The solid curve in the figure gives the total production cross section as reconstructed from the asymptotic $\pi^0\gamma$ yield without any restriction on the $\pi^0\gamma$ invariant mass. Also shown is the contribution from actual \w mesons produced in exclusive photoproduction processes (dotted curve) as well as the contribution of all \w mesons produced (dashed curve). One sees that the latter curve lies below the total reconstructed yield. This is due to rescattering of the $\pi^0$ on nucleons which destroys the $\pi^0\gamma$ correlation. For higher incident photon energies an additional combinatorial background contributes since the final state might contain more than one $\pi^0$, therefore making the reconstruction of the \w ambiguous. This background cannot be suppressed by applying an exclusivity measure that confines the primary production process to $\gamma N\rightarrow\omega N$ since one would lose up to 90\% of the slow \w mesons which are produced in inclusive photoproduction processes and which are the dominant source of information about the \w in-medium properties.

\begin{figure}
\begin{center}
\includegraphics[scale=1.5]{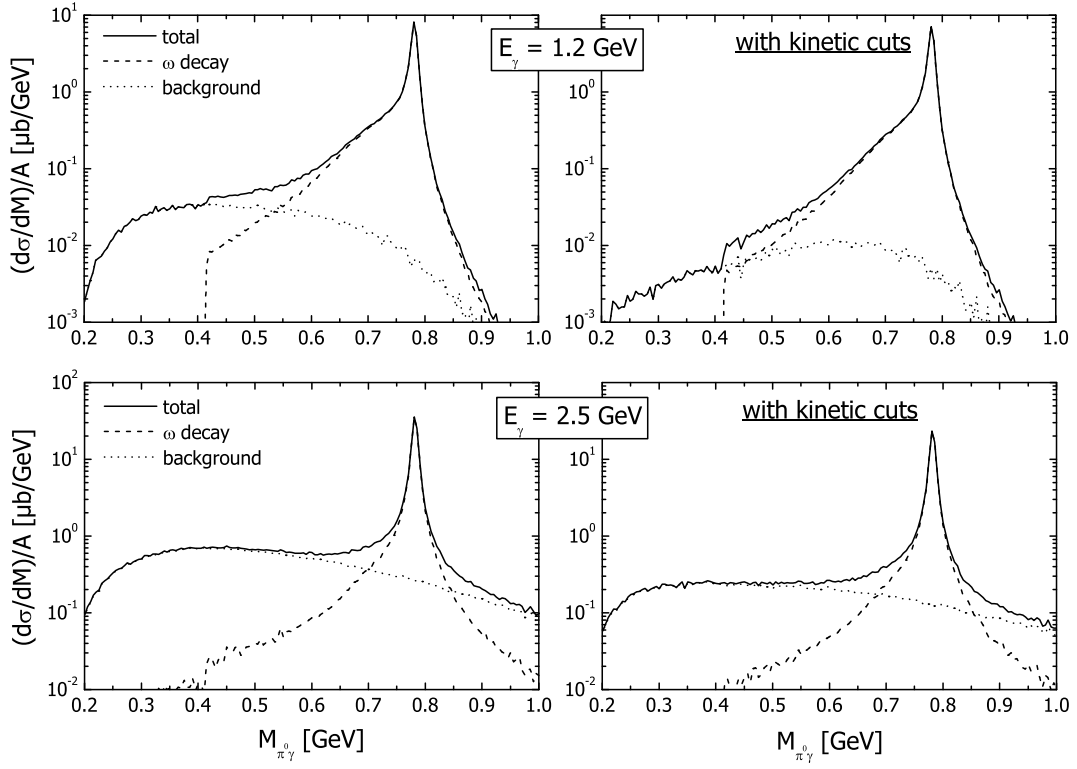}
\caption{Mass differential cross section for the reaction $\gamma Nb\rightarrow\pi^0\gamma X$ at 1.2 GeV (upper panels) and 2.5 GeV (lower panels) incident photon energy, with (right) and without (left) applying cuts on the kinetic variables of the final state particles. See text for details.}
\label{figure06}
\end{center}
\end{figure}

As already pointed out in Ref. \cite{Cassing:2001}, the background from rescattering can be suppressed quite successfully by applying a lower cut on the kinetic energy of the outgoing $\pi^0$ since these pions in average lose a great amount of their kinetic energy in the collisions. We therefore require that the kinetic energy of the final state $\pi^0$ should be larger than 150 MeV. At the higher beam energies we apply an additional cut on the out-of-plane angle, which is the angle between the $\pi^0$ direction and the initial and final state $\gamma\gamma$ plane. If an \w would be produced with a momentum parallel or anti-parallel to the direction of the incoming photon from a nucleon at rest, the sine of the out-of-plane angle would equal to zero. A cut of $\sin{\phi}\le 0.5$ therefore not only cuts some background away but also triggers on \w mesons produced parallel or antiparallel to the incoming photon direction, corresponding to \w mesons with the highest or lowest kinetic energies. In Fig. \ref{figure06} we demonstrate the effect of the kinematic cuts. To suppress an additional combinatorial background stemming from the production of $\pi^0\pi^0$ pairs by not detecting one of the photons from the $\pi^0$ decay, the authors of Ref. \cite{Cassing:2001} proposed an additional cut on the energy of the outgoing photon $E_{\gamma}>200$ MeV. This cut essentially suppresses the entire background coming from these processes and, therefore, has been applied to all the observables we are going to discuss in the following.  

Only $\pi^0\gamma$ pairs originating from \w decays inside the nucleus carry information on the \w in-medium properties. Therefore the kinematical condition arises that the decay length of the \w, $L=\gamma\beta\tau=p/(m\Gamma)$, is in the same order as the nuclear radius of the target material. Hence, for a set of ordinary target nuclei the maximum value of the \w momentum should be in the range of $100-200$ MeV. Apparently \w mesons produced from photonuclear reactions gain much higher momenta as can be seen in Fig. \ref{figure05}. To suppress a large superimposed signal from \w decays in vacuum it might be useful to apply an upper momentum cutoff to the three-momentum of the $\pi^0\gamma$ pair.

\begin{figure}
\begin{center}
\includegraphics[scale=1.4]{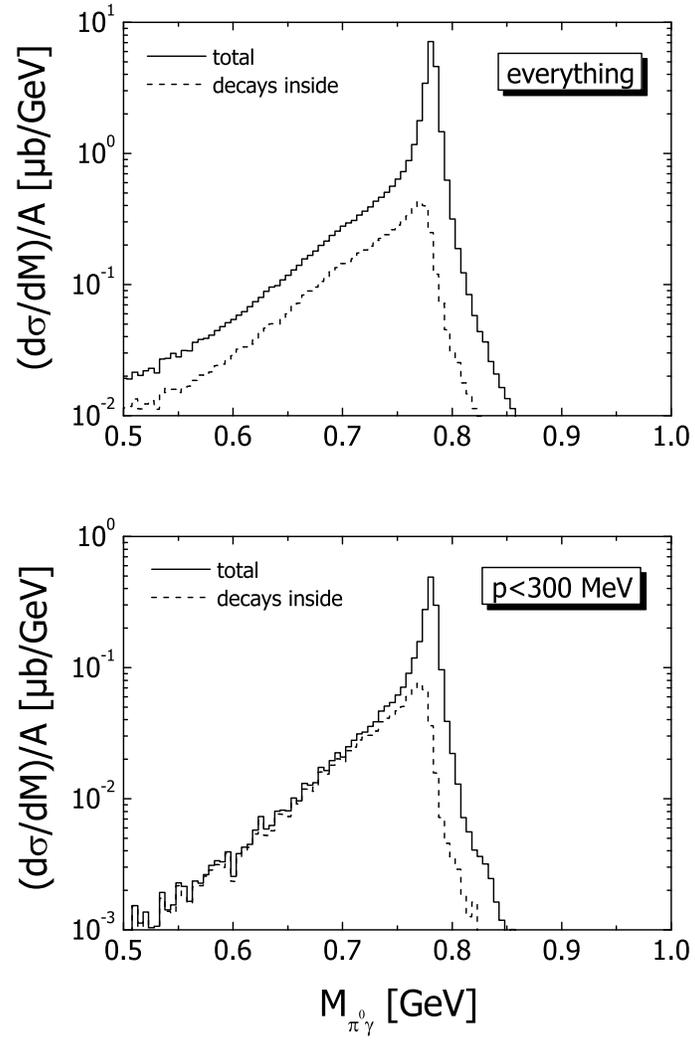}
\caption{Invariant mass distribution at 1.2 GeV incident photon energy. Indicated by the dashed line is the contribution from \w decays inside the nucleus. Upper panel: without momentum cut, lower panel: with a momentum cut of 300 MeV.}
\label{figure07}
\end{center}
\end{figure}

\begin{figure}
\begin{center}
\includegraphics[scale=1.5]{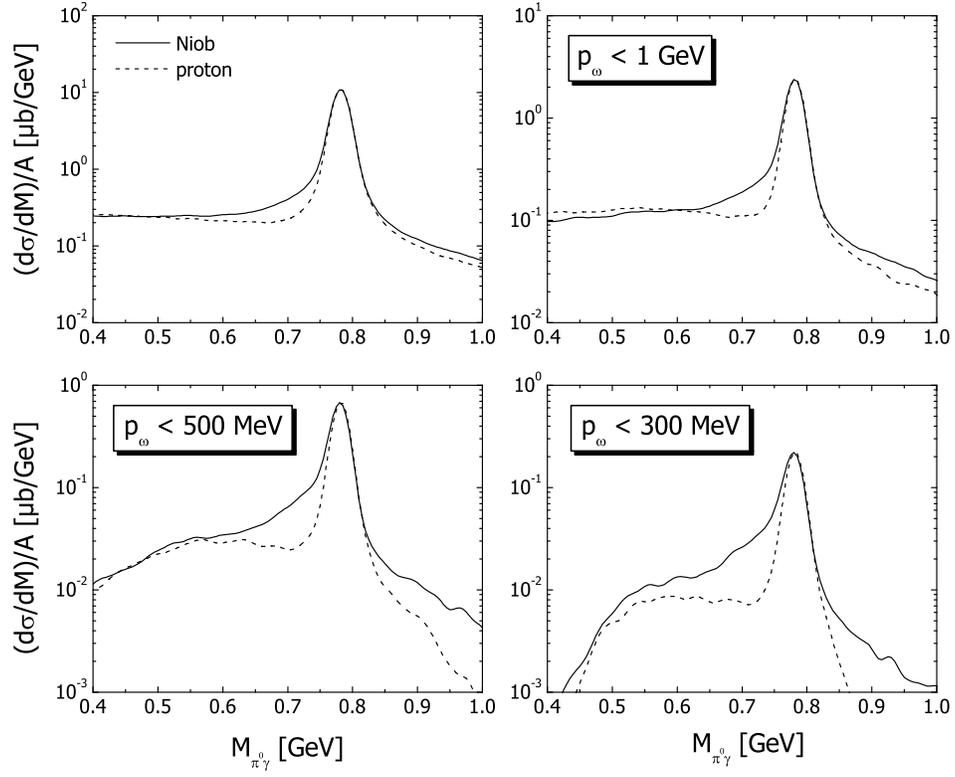}
\caption{Comparison of the invariant mass distribution obtained from proton and Niobium targets for several values of the momentum cutoff at 2.5 GeV incident photon energy. The folding resolution is $\Delta m=10$ MeV. The cross sections on the proton are normalized to the cross section on Niobium at the \w pole mass $m=782$ MeV.}
\label{figure08}
\end{center}
\end{figure}

In Fig. \ref{figure07} we demonstrate the effect of a momentum cutoff of 300 MeV on the invariant mass distribution for an incident beam energy of 1.2 GeV. The following results involve medium modifications of the \w meson as given by equations (\ref{width}) and (\ref{shift}). Applying the momentum cut, the ratio of in-medium to vacuum signal increases. However, one also loses almost one order of magnitude of the total cross section. We have checked that at this beam energy a momentum cutoff yields only a  weak improvement of the signal for the \w in-medium changes and, therefore, we present the results at 1.2 GeV photon energy without a cutoff in the \w three momentum. At an incident beam energy of 2.5 GeV a cut on the \w momentum becomes more essential. Note that in this momentum range the exclusive cross section amounts only to a very small part of the total cross section (cf. Fig. \ref{figure05}). In Fig. \ref{figure08} we compare the results on Niobium to those on the proton without a momentum cut and with cutoff values of 1 GeV, 500 MeV, and 300 MeV. All calculated spectra we are going to show in the following are folded with an experimental mass resolution of 10 MeV, which, however, might be hard to achieve in experiment. The cross sections on the proton are normalized to the cross section labeled "Niob" at the \w pole mass $m=782$ MeV (see figure captions). With a decreasing cutoff we find an increasing modification of the invariant mass spectrum on the nucleus on the low mass side of the \w peak as compared to the spectrum obtained from the proton due to increasing average decay densities. A drawback of the clear signal which we obtain with a cutoff of 300 MeV is the small cross section.

\begin{figure}
\begin{center}
\includegraphics[scale=1.4]{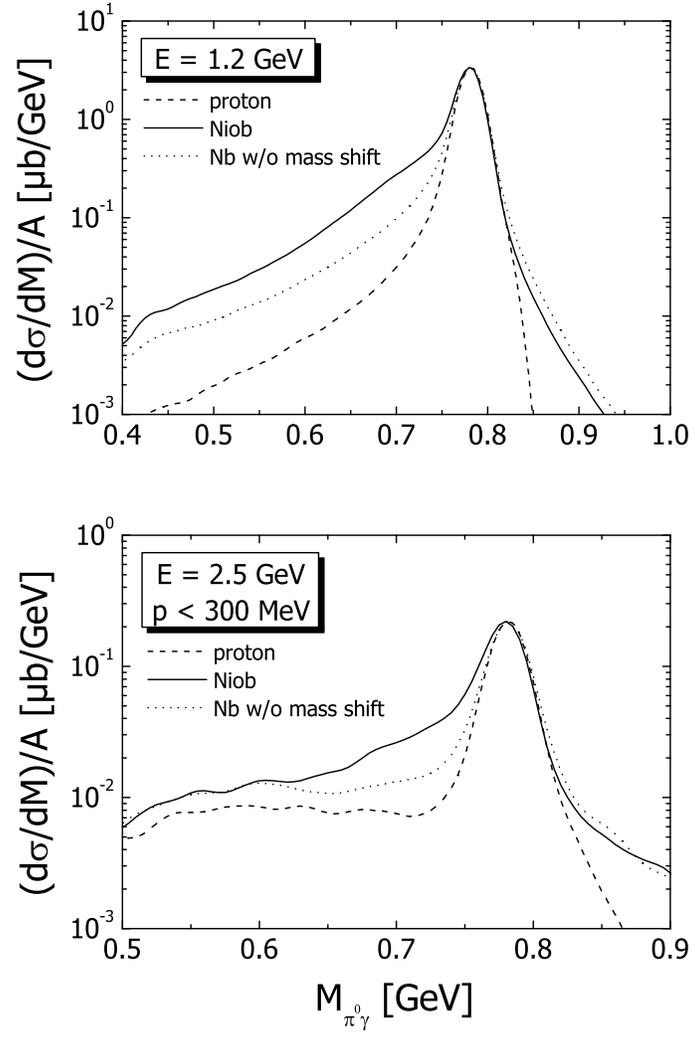}
\caption{Mass differential cross section at 1.2 and 2.5 GeV photon energy. Comparison of cross sections on proton, Niobium, and Niobium without a mass shift of the $\omega$. The dotted curve contains collisional broadening only. The cross sections on the proton and on Niobium without mass shift are normalized to the Niobium cross section including the dropping \w mass.}
\label{figure09}
\end{center}
\end{figure}

In Fig. \ref{figure09} we show our results for both incident photon energies. In order to facilitate the comparison of the results we normalize all curves to the one labelled "Niob" , i. e. to the one containing the in-medium changes. We will show later in Fig. \ref{figure12}, that for the nuclear targets this normalization amounts only to a small change in the peak value. To get a handle on which part of the signal can be attributed to a shift of the \w pole mass and which part comes from collisional broadening, we present an additional curve (dotted) which is calculated without the dropping \w mass. For both incident beam energies only a part of the accumulated spectral strength below the \w pole mass is due to the \w in-medium broadening. For 1.2 GeV photon energy about half of the in-medium signal -- and even more for 2.5 GeV photon energy -- originates from the downward shift of the \w pole mass and, therefore, reveals a non-trivial in-medium change of the isoscalar spectral density. Qualitatively our results shown in Fig. \ref{figure09} look very similar to what has been obtained in Ref. \cite{Cassing:2001}. The somewhat more pronounced structure in the low mass part of the \w mass spectrum shown in Fig. 8 of Ref. \cite{Cassing:2001} might possibly be explained by the use of a constant collision width for the \w as opposed to the density and momentum dependent width in our approach. Overall the agreement of the two independent approaches can be seen as a confirmation of the conclusions drawn in Ref. \cite{Cassing:2001} and the present work.

\begin{figure}
\begin{center}
\includegraphics[scale=1.4]{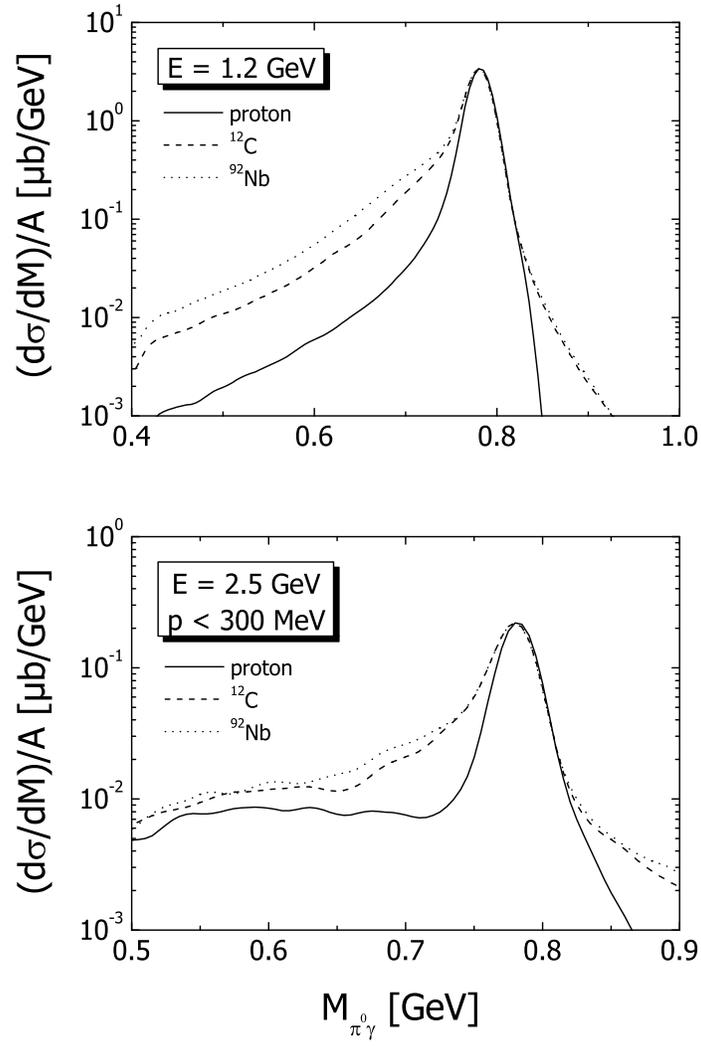}
\caption{Comparison of mass differential cross sections on proton, Carbon, and Niobium targets. The cross sections on Carbon and on the proton are normalized to the Niobium cross section at the \w pole mass.}
\label{figure10}
\end{center}
\end{figure}

In Fig. \ref{figure10} we compare the invariant mass spectra on the proton, Carbon ($A=12$) and Niobium ($A=92$). We find only a very weak dependence of the medium signal on the atomic mass of the considered nuclei. The reason for this is that most of the observed  $\pi^0\gamma$ pairs stem from the nuclear surface because of the strong $\pi^0$ absorption in the FSI. Therefore the visible fraction of \w mesons decaying inside the nucleus cannot be increased by further increasing the size of the target nucleus. We, therefore, restrict ourselves to the discussion of the results on Niobium, which is one of the considered target materials of the TAPS/Crystal Barrel experiment, expecting qualitatively very similar results for all other nuclei.

\begin{figure}
\begin{center}
\includegraphics[scale=1.4]{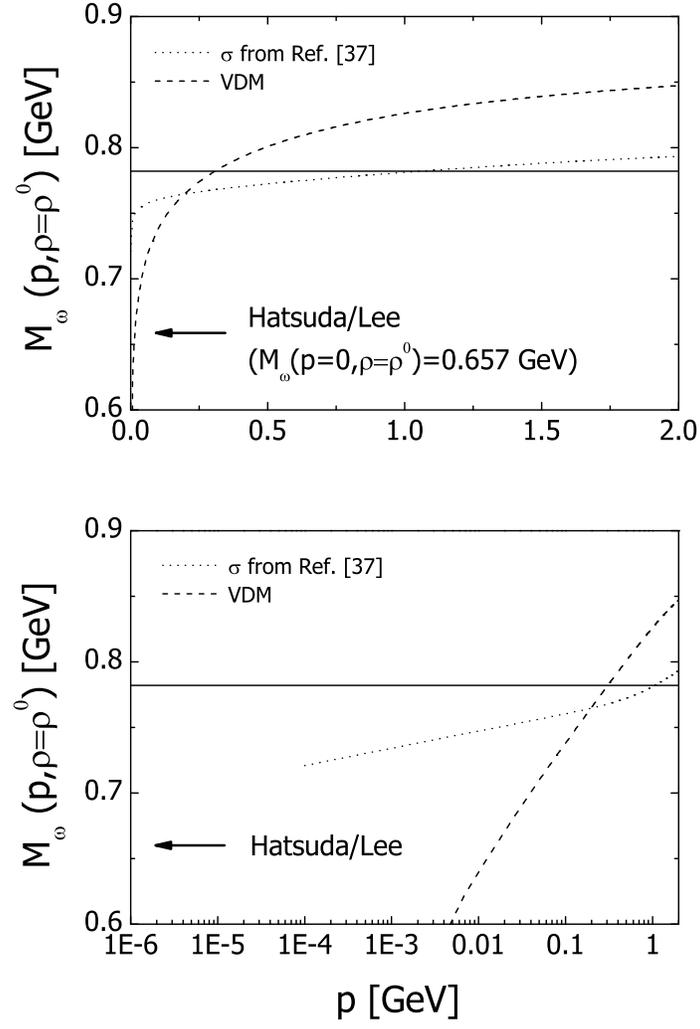}
\caption{Effective \w mass as function of the \w momentum at $\rho^0$ calculated with the total $\omega N$ cross section of Ref. \cite{Lykasov:1998} and a cross section obtained via VMD.}
\label{figure11}
\end{center}
\end{figure}

\begin{figure}
\begin{center}
\includegraphics[scale=1.4]{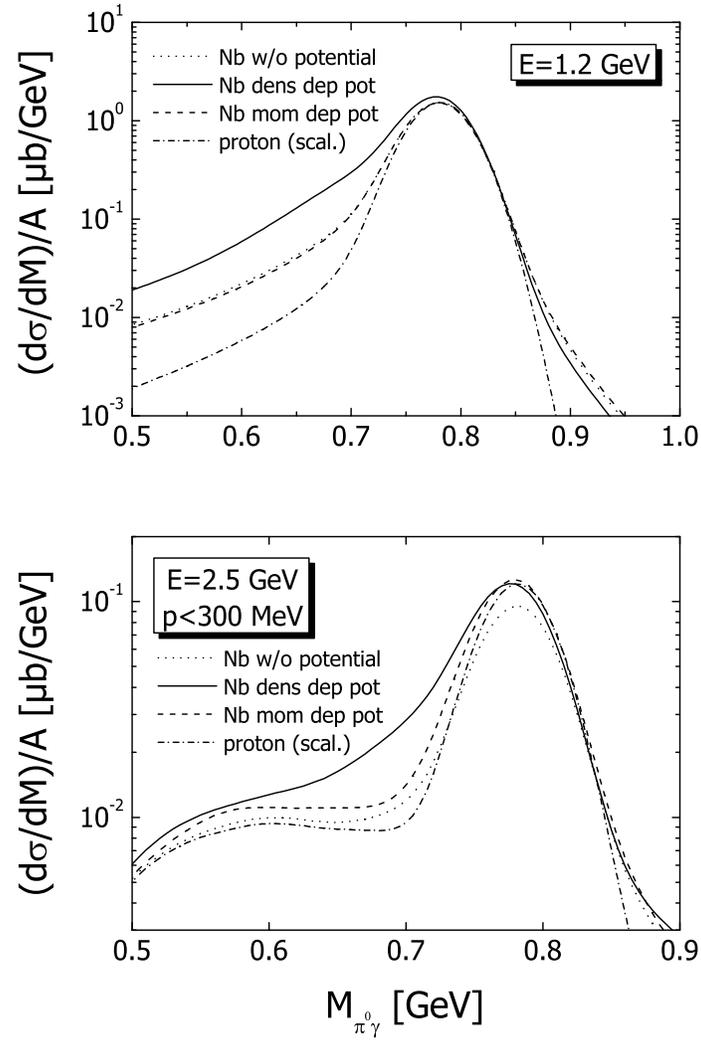}
\caption{Invariant mass spectra for Niobium targets at 1.2 and 2.5 GeV incident photon energy with a mass resolution of 25 MeV. The proton cross section is normalized to the cross section on Niobium obtained with the momentum dependent \w potential.}
\label{figure12}
\end{center}
\end{figure}

Up to now we modified the in-medium mass of the \w according to Eq. (\ref{shift}). However, this expression is valid for \w mesons at rest, whereas the momenta of the \w mesons produced in the photon nucleus reaction deviate considerably from zero, see Fig. \ref{figure05}. In lowest order in the density the effective \w mass in-medium is connected to the real part of the $\omega N$ forward scattering amplitude by the following relation:
\bea
{m^*}^2=m^2-4\pi Re f_{\omega N}(\nu,\theta=0)\rho_N.
\eea
The real part of the forward scattering amplitude $f_{\omega N}(\nu,\theta=0)$ can be calculated from its imaginary part via a subtracted dispersion relation \cite{BjorkenDrell}
\bea
Re f(\nu,0)=Re f(\nu_0,0)+\frac{2(\nu^2-\nu_0^2)}{\pi} \mathcal{P}\!\!\int\limits_{\nu_{\mathrm{min}}}^{\infty}\frac{d\nu'\nu'Im f(\nu',0)}{({\nu'}^2-\nu_0^2)({\nu'}^2-\nu^2)},
\eea
where $\nu_{\mathrm{min}}$ is the threshold energy and $\nu_0$ is a subtraction point. The imaginary part of the forward scattering amplitude is connected with the $\omega N$ total cross section via the optical theorem
\bea
Im f_{\omega N}(\nu,0)=\frac{p}{4\pi}~\sigma_{\omega N}^{\mathrm{tot}}(\nu),
\eea
where $p$ is the \w momentum in the nucleon rest frame. In Fig. \ref{figure11} we show the effective \w mass as a function of the \w momentum in the nuclear rest frame. For the ratio of the real to the imaginary part of the forward scattering amplitude at the subtraction point $\nu_0=4$ GeV we used a value of $\alpha(\nu_0)=Re f(\nu_0,0)/Im f(\nu_0,0)=-0.1$, which on one hand is motivated by measurements of the $\rho N$ and $\phi N$ forward scattering amplitudes \cite{Alvensleben:1970,Alvensleben:1971,Biggs:1971} and on the other hand yields a result that is consistent with our previous choice for the \w mass in-medium at zero momentum, see Fig. \ref{figure11}. Besides the calculation employing the total $\omega N$ cross section of Ref. \cite{Lykasov:1998} (dotted curve), which we are also using for $\omega N$ collisions throughout the FSI, we also show a curve for which the total $\omega N$ cross section has been obtained via VMD from the differential \w photoproduction cross section at $t=0$ with a value of $\alpha(\nu_0)=-0.3$ (dashed curve) as determined for tne $\rho N$ forward scattering amplitude in Ref. \cite{Biggs:1971}. The VMD result shows a much stronger momentum dependence of the effective \w mass. However, the dispersion integral is dominantly sensitive to the low energy part of the cross section and one does not expect that the VMD model is valid for these energies. A similar result, i. e. a rising vector meson mass with increasing vector meson momentum, has been found in a more refined study in Ref. \cite{Eletsky:1996} for the $\rho$ meson.

In order to include this momentum dependence in our calculations we adopt the following parameterization:
\bea
\label{momdep}
m_{\omega}^*=m_0\left(1-\beta\left (1-\gamma \left[\frac{p}{\mathrm{GeV}} \right]^{\delta}\right)\frac{\rho_N}{\rho_0}\right)
\eea
with the constants $\beta=0.16$, $\gamma=1.0$, and $\delta=0.11$. This parameterization reproduces the \w mass at zero momentum of Ref. \cite{Hatsuda} and then goes over into the result for the momentum dependent \w mass employing the $\omega N$ total cross section of Ref. \cite{Lykasov:1998}. Our calculation of the effective momentum dependent \w mass meant to give an educated guess for the momentum dependence of the in-medium \w mass in order to explore its consequences on a measurement of the $\pi^0\gamma$ invariant mass spectrum.

In Fig. \ref{figure12} we show the results obtained with the momentum dependent \w mass (Eq. (\ref{momdep})) together with the calculations using the purely density dependent in-medium mass (Eq. (\ref{shift})) and the calculations without any vector meson potential in comparison to the results for $\pi^0\gamma$ production off the proton. These spectra are folded with a mass resolution of 25 MeV, which is a more realistic value in view of the experimental determination of the $\pi^0\gamma$ invariant mass spectrum. For both incident beam energies the results obtained with the momentum dependent potential are very similar to the results with the strong \w potential set to zero. This is due to the fact that at momenta of a few hundred MeV the mass shift becomes very small. Those momenta receive the strongest weight due to the steep increase of the cross section with $p$, cf. Fig. \ref{figure05}. Hence, any lowering of the mass of a slow \w inside the nucleus could only be detected by a significantly smaller momentum cut, which, however, might experimentally not be feasible due to the extremely small cross section. On the other hand, the measurement of such an invariant mass spectrum as plotted in Fig. \ref{figure12} would not a priori rule out a mass shift of the \w meson at rest.

\section{Summary}\label{summary}

The photoproduction of $\pi^0\gamma$ pairs has been studied in the kinematical region of an ongoing experiment with the Crystal Barrel and TAPS detectors at ELSA. Exploiting a semi-classical transport model we have studied the feasibility to learn about possible in-medium changes of the \w meson by comparing the invariant mass spectra obtained by photoproduction on the proton and complex nuclei. Different assumptions on the \w in-medium mass and width have been adopted in the transport simulations in order to explore the consequences on the experimental determination of the \w in-medium properties via $\pi^0\gamma$ photoproduction off nuclei.

Despite the large average momenta of the \w mesons produced in high energy photon-nucleus reactions it is still possible to observe a signal stemming from \w in-medium decays, eventually applying a cutoff on the three momentum of the reconstructed \w mesons. Slow \w mesons, which are the main source of information about the \w in-medium changes, are at large photon energies produced
dominantly by inclusive photoproduction processes, which have been included in the transport simulations by means of the hadronic event generator FRITIOF. For incident beam energies of 1.2 and 2.5 GeV a low-mass tail of the \w meson is visible in the $\pi^0\gamma$ mass spectrum, revealing a non-trivial downward shift of the \w mass at nuclear densities. As also pointed out in Ref. \cite{Cassing:2001}, the background from rescattering of the $\pi^0$ in the medium can be suppressed successfully by applying kinematic cuts on the observables.

Due to the large absorption cross section of the $\pi^0$ the obtained signal is dominantly sensitive to the nuclear surface. The average decay density that can be seen in the detector can hardly be increased by increasing the size of the nucleus. We, therefore, find only a very weak dependence of the shape of the low-mass tail of the \w on the atomic mass of the considered nucleus.

A momentum dependence of the strong \w potential has been estimated by means of a subtracted dispersion relation. Adopting this potential into the transport simulations, we found results which are very similar to the scenario involving a non-dropping \w mass. A signal of the lowering mass could only be obtained by decreasing the momentum cutoff well below the value where the in-medium mass becomes small. This, of course, comes of the prize of significantly reduced counting rates. However, measurements with varying momentum cuts may then help to disentangle the density- and momentum dependence of the in-medium self-energy of the \w meson.

In summary, it has been shown that the reaction $\gamma A\rightarrow\pi^0\gamma X$ can be exploited to gain information about the \w spectral distribution in nuclei. The experimental determination at energies slightly above threshold as well as at higher photon energies could shed some light not only on the existence of a downward shift of the \w mass at finite baryon density but also might provide restrictions on its momentum dependence, what up to the present is an entirely unsettled issue.

\acknowledgments{The authors acknowledge stimulating discussions with D. Trnka and V. Metag. The authors would like to thank DFG and BMBF for financial support.}

\pagebreak

\end{document}